\documentclass[conference]{IEEEtran}
\IEEEoverridecommandlockouts
\usepackage{cite}
\usepackage{amsmath,amssymb,amsfonts}
\usepackage{algorithmic}
\usepackage{graphicx}
\usepackage{textcomp}
\usepackage{xcolor}
\usepackage{url}
\def\BibTeX{{\rm B\kern-.05em{\sc i\kern-.025em b}\kern-.08em
    T\kern-.1667em\lower.7ex\hbox{E}\kern-.125emX}}
    
\begin{document}

\title{Photon: A Cross Platform P2P Data Transfer Application\\
}

\author{%
\small
  \begin{tabular}[t]{c}
    \textbf{Abhilash Shreedhar Hegde} \\
    Undergraduate\\
   Dept of Computer Science \& Engineering \\
    The National Institute of Engineering \\
    Mysuru, Karnataka, India \\
    \texttt{4ni19cs003\_a@nie.ac.in}
  \end{tabular} \quad
  \begin{tabular}[t]{c}
    \textbf{Amruta Narayana Hegde} \\
     Undergraduate\\
   Dept of Computer Science \& Engineering \\
    The National Institute of Engineering \\
    Mysuru, Karnataka, India \\
    \texttt{4ni19cs020\_a@nie.ac.in}
  \end{tabular} \quad
  \begin{tabular}[t]{c}
    \textbf{Adeep Krishna Keelar} \\
     Undergraduate\\
     Dept of Computer Science \& Engineering \\
    The National Institute of Engineering \\
    Mysuru, Karnataka, India \\
    \texttt{4ni19cs007\_a@nie.ac.in}
  \end{tabular} \\[2ex]
  \begin{tabular}[t]{c}
    \textbf{Ananya Mathur} \\
     Undergraduate\\
    Dept of Computer Science \& Engineering \\
    The National Institute of Engineering \\
    Mysuru, Karnataka, India \\
    \texttt{4ni19cs022\_a@nie.ac.in}
  \end{tabular} \quad
  \begin{tabular}[t]{c}
    \textbf{C. Vidya Raj} \\
    Professor \& Dean \\
    Dept of Computer Science \& Engineering\\
     The National Institute of Engineering \\
    Mysuru, Karnataka, India \\
    \texttt{vidyarajc@nie.ac.in}
  \end{tabular}
}
\date{\today}

\maketitle

\begin{abstract}
Modern computing requires efficient and dependable data transport. Current solutions like Bluetooth, SMS (Short Message Service), and Email have their restrictions on efficiency, file size, compatibility, and cost. In order to facilitate direct communication and resource sharing amongst linked devices, this research study offers a cross-platform peer-to-peer (P2P) data transmission solution that takes advantage of P2P networks' features. 

The system enables cost-effective and high-performance data transport by using the compute, storage, and network resources of the participating devices. Simple file sharing, adaptability, dependability, and high performance are some of the important benefits. The examination of the suggested solution is presented in this paper and includes discussion of the P2P architecture, data transfer mechanisms, performance assessment, implementation issues, security concerns, and the potential difficulties that needs to be addressed.

The research intends to validate the efficacy and potential of the suggested cross-platform P2P data transfer solution, delivering better efficiency and dependability for users across various platforms, through practical investigations and comparisons with existing approaches.
\end{abstract}


\section{\textbf{Introduction}}
Data transfer is the application of computing techniques to transmit electronic or analog data from one device to another. The transferred data may be of any type, size, context, and nature and it can be accomplished through network-less modes, from copying data to an external device and then copying from that device to another. However, this is a time-consuming task and ,even, an irritating job, in case of multiple changes and transferring of the same file is done. While some wireless and communication features enable data transfer smoothly, these being Bluetooth, SMS (Short Message Service), and Email, they all possess several flaws. Bluetooth does provide good efficiency for very nearby devices and small files, however, takes a large amount of time for larger files and is sometimes incompatible with devices that are of lower quality or branding. Data transfer through SMS is impractical considering its limitations in size and cost, while Emails, too, have similar flaws. \\

A peer-to-peer network is an information technology infrastructure that allows two or more computer systems to connect and share resources without requiring a separate server or server network. Unlike the client-server architecture, there is no central server for processing requests and sending responses, for the peers directly interact with one another. Each computer in a P2P network provides resources to the network and consumes resources that the network provides. Resources such as files, printers, storage, bandwidth, and processing power can be shared between various computers in that network. Advantages of a P2P Network, within the local area, include easy file sharing, reduced costs, adaptability, reliability, high performance, and efficiency.\\

With the existing options with some flaws or with the application in need having to face some of the flaws with either poor access to the internet and or previous solutions being non-functional, our proposal aims to address the above-mentioned with a solution that can tackle the issues on a wide variety of platforms with ease. Our solution aims to be cross-platform, thus able to run on any environment and provide the best performance. 

\section{\textbf{Literature Survey}}

\textbf{Turbo Share} – File sharing application developed by \textit{Suraj Bhul et al.}, [5], was based on Peer-to-Peer data transfer technology. The app supports features like multiple file transfer, viewing file sharing history, average data transfer speed, etc. But it lacks cross-platform availability as it supports only android devices. \\ 

A paper published by \textit{Paulo R. M. de Andrade, et al.},[2], describes how cross-platform applications work and a comparison of the same with native applications. It draws contrasting features between hybrid, native, and web apps. It shows how operating systems have evolved, what are the needs for cross-platform apps, and the richness offered by them when compared to native apps. It also gives statistical insight into the usage of several types of applications (native, cross-platform, and web apps). \\

A paper published by \textit{Yoonsik Cheon} and team [3], describes their working experience on converting their natively built Android application to cross-platform by means of the Flutter framework. The paper describes their application that was built on a Java platform and how it was to convert the application using the Flutter framework to make it cross-platform.  \\

A paper by \textit{Mr. D. S. Thosar} and team, [4], reviews the aspects of file sharing using a local area network in a peer-to-peer network. The paper speaks of the various drawbacks that Bluetooth, SMS (Short Message Service) and Internet based transfer of data ie files face and how a minimal set client to client-based transfer technology can help to transfer files.  \\

A paper by \textit{Wenhao Wu} [1] describes cross-platform app development using React Native and Flutter. React Native is a JavaScript framework that lets developers build apps using JavaScript. It uses a design principle like React. It uses declarative approach to design the User Interface (UI). React Native uses the same fundamental user interface as the building blocks for native android and iOS apps. The same paper also describes about the '\textbf{Flutter}' framework which unlike React Native, does not use JavaScript but rather uses a newly developed language called '\textbf{Dart}'. Dart was developed and is maintained by Google. It provides features like cross- platform support, rich User Interface features and hot-reload. 

\section{\textbf{Proposed Solution: Photon}}
Photon is a cross-platform file-transfer application built using the Flutter framework. It uses HTTP to transfer files between devices. One can transfer files between devices that run Photon. (No Wi-Fi router is required; one can use hotspot). Some of the features of photon are,\\

\begin{itemize} 
    \item Cross-platform support - For instance, one can transfer files between Android and Windows, GNU/Linux based, Macintosh based systems and vice-versa. 
    \item Transfer multiple files - One can pick any number of files and transfer them efficiently. 
    \item Smooth User Interface - In contrast to the strictly bound design paradigms such as Model-View-View Model (MVVM), Model-View-Presenter (MVP), Clean Architecture and such, the application uses the "\textbf{Material You}" design paradigm, an evolution of the Material Design language. Introduced by Google in 2014, it focuses on customization, personalization, and expressing individuality within the user interface.  
    \item Minimal internet usage - Transfer of data over the internet, while is efficient and reliable given the factors being favourable, has the drawbacks of being expensive, and in case of unfavourable conditions might suffer from latency. However, in this application, data transfer is facilitated by using Wireless Fidelity (Wi-Fi) Direct Technology or P2P (Peer to Peer), allowing devices to connect directly with each other via Wi-Fi without an intermediate access point. This type of method provides faster transfer speeds when compared to other technologies such as Bluetooth. 
    \item Security - Uses a secure secret code generation for authentication (internally). Even though the files are streamed at local area network, these files cannot be downloaded/received without using the application. Thus, no external client like a browser can get the transferred files using an URL as the secret code is associated with the URL. It will be regenerated for every session. 
    \item Supports high-speed data transfer - Photon is capable of transferring files at a remarkably high rate, depending upon the Wi-Fi bandwidth however, no internet connection required. 
\end{itemize}

\subsection{\textbf{Architecture of Photon}}\label{AA}
Photon was built using the Flutter framework, designed to provide a fast, efficient, and reactive application experience. It is widget-based approach and reactive programming model, combined with the built-in navigation system and rich set of tools, making it highly feasible for developing high-quality, responsive and performance applications.  \\

\begin{itemize} 
    \item Widgets - Being the building blocks of Flutter UI (User Interface), from buttons to layout elements, it describes its UI elements and behaviour. It consists of two types - StatelessWidgets representing those widgets which cannot change its state and StatefulWidgets that can be updated and rebuilt. 
    \item Composition - Flutter follows a composition model, complex UI elements are built by combining multiple widgets which can be either nested or combined in various methods to get the desired interface. 
    \item Reactive Framework - Changes to the user interface are triggered by the changes in the application state, defining the reactive behaviour. This approach helps to keep the user interface coordinated with the underlying states.  
    \item Hot Reload - Flutter allows developers to make changes in the code and see the changes immediately without restarting the application, speeding the development process by providing a fast feedback loop facilitating iterative development.  
    \item Rendering - Flutter bypasses the native UI components of the underlying platform and uses its own rendering engine called \textit{Skia} which renders its UI to the native canvas, resulting in a highly customizable and performant UI experience.  \\
\end{itemize}

Photon was built using the Model-View-Controller (MVC) architecture pattern, the Model being the data structures that provides necessary data, the View being the interface that the user gets to see and the Controller being the intermediate link between the View and the Model, responsible for handling user input, updating the model and view appropriately. The separation between the three layers makes it easy to debug and helps to build stable systems with the clear separation between the application and the business logic. The changes in the View will not affect the Model, making it easier to make changes without breaking the application. In the case of Photon being built with the Flutter framework, multiple screens and views using the same logic reduces code duplicity, increasing code reusability.

\subsection{\textbf{Working of Photon}}
Photon works between two interacting devices, one being a Sender of some data and the other being the Receiver of the shared Data. The Sender object initially upon starting the Photon application gets to pick two options (Send and Receive) and upon choosing Send, the sender first picks the data (be files/pictures/videos/apk files) that sender wishes to send from the device (be mobile/desktop) and starts the server. Simultaneously, the other object the Receiver upon choosing the Receive option first must discover the Sender within the same network and upon discovery, sends a Permission request to the Sender who can either deny or accept permission to start the session between the two. A session is generated between the two objects where the Receiver gets the file index from the sender who also transfers the files to the Receiver. After receiving the file, the Receiver can terminate from the session and end the session, while the Sender too ends the Session with the Receiver by stopping the server.  \\

Some of the functionalities are described. 

\begin{itemize} 
    \item Writing services enables end to end functionality of the application.  
    \item To facilitate the file picking mechanism, the application should be able to pick files from the device which is provided by the \textit{file\_picker} package to pick files from the local storage.  
    \item To implement the file transfer mechanism, the application uses Hyper Text Transfer Protocol (HTTP). When Peer 1 sends the files to Peer2, it sets up a HTTP-server locally which is received by Peer2 when it makes a HTTP-GET request. To improve the efficiency, the application does not load the whole file into the memory streams the file into bytes of data, hence the device never runs out of memory.  
    \item Error Handling is an important feature that helps to handle a lot of unknown scenarios that prevent the smooth functioning of the application, Photon is coded with several try-catch cases.  
\end{itemize} 

\begin{figure} 
    \centering 
    \includegraphics[width=1\linewidth]{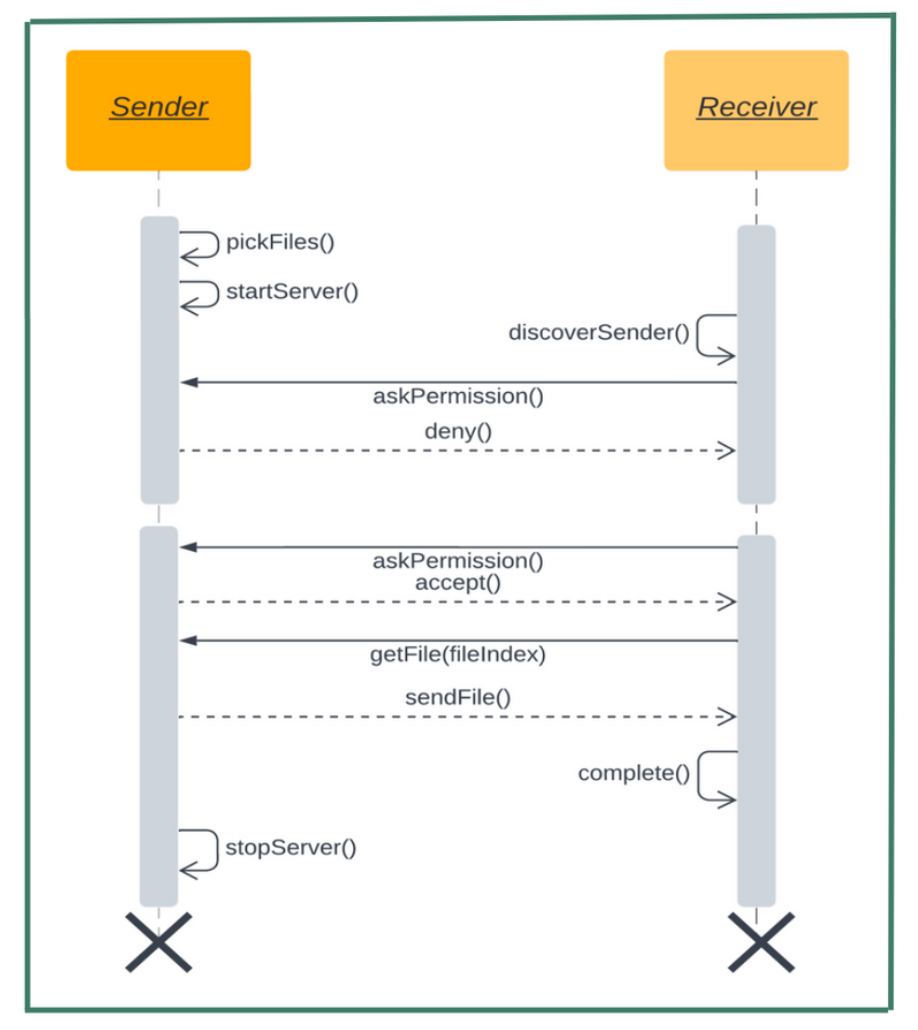} 
    \small Figure 1. {Sequence diagram showing interaction between peers} 
    \label{fig:} 
\end{figure} 

\subsection{\textbf{Performance Check of Photon}} 

To check the performance of Photon, the application was tested against a variety of devices ranging from Mobile-to-Mobile transfers, Mobile to Desktop Transfers and Desktop to Desktop Transfers. Mobile environments include Android and iOS while Desktop environments include Windows (Windows 10, Windows 11), GNU/Linux (Ubuntu 18.10, Ubuntu 20.04) and Macintosh OS (Operating System) (Ventura). The results are tabulated to various files of sizes ranging from few megabytes to 5 gigabytes are displayed in the table I. 

\begin{table}[htbp]
\caption{File size vs Time taken}
\begin{center}
\begin{tabular}{|c|c|}
\hline
\cline{2-2} 
 \textbf{\textit{File size in MB}}& \textbf{\textit{Time taken in seconds}}\\
\hline
    1 & 0.039 \\
    10 & 0.4 \\
    20 & 1 \\
    50 & 2 \\
    100 & 5 \\
    1000 & 62 \\
    5000 & 314\\
    
\hline
\multicolumn{2}{l}{$^{\mathrm{a}}$Between macOS Ventura and Android 13.}
\end{tabular}
\label{tab1}
\end{center}
\end{table}

\begin{table}[htbp]
\caption{Transferring data between different devices}
\begin{center}
\begin{tabular}{|c|c|c|}
\hline
\cline{3-3}
 \textbf{\textit{System Type}}& \textbf{\textit{PDF File}}& \textbf{\textit{Video File}}\\
\hline
    Android Samsung & 5s & 17.20s \\
    iOS iPhone & 4s & 16.80s \\
    Dell Inspirion 3410 AMD Ryzen 5 Win 10 & 4.5s & 16.4s \\
    Dell Inspirion 3000 Intel i5 5th Gen (Linux) & 8s & 24.4s \\
    MacBook Air & 5s & 11.4s \\
\hline
\hline
\multicolumn{2}{l}{$^{\mathrm{a}}$PDF size - 100MB and High Quality Video - 300MB}
\end{tabular}
\end{center}
\end{table}

\subsection{\textbf{Testing}}

One key aspect of system testing for a Flutter app is testing for compatibility across different platforms. This means testing the app on a range of devices, such as smartphones and tablets, with different screen sizes and resolutions, as well as testing the app on different operating systems such as Android, GNU/Linux, Windows and macOS. This helps ensure that the app functions correctly on all platforms and that users have a consistent experience across devices. For testing software, various test strategies are to be used such as unit testing, integration testing, system testing etc. The project was put through intensive testing methods to ensure efficiency and accuracy.

\begin{table}
\caption{Test cases}
    \centering
    \includegraphics[width=1\linewidth]{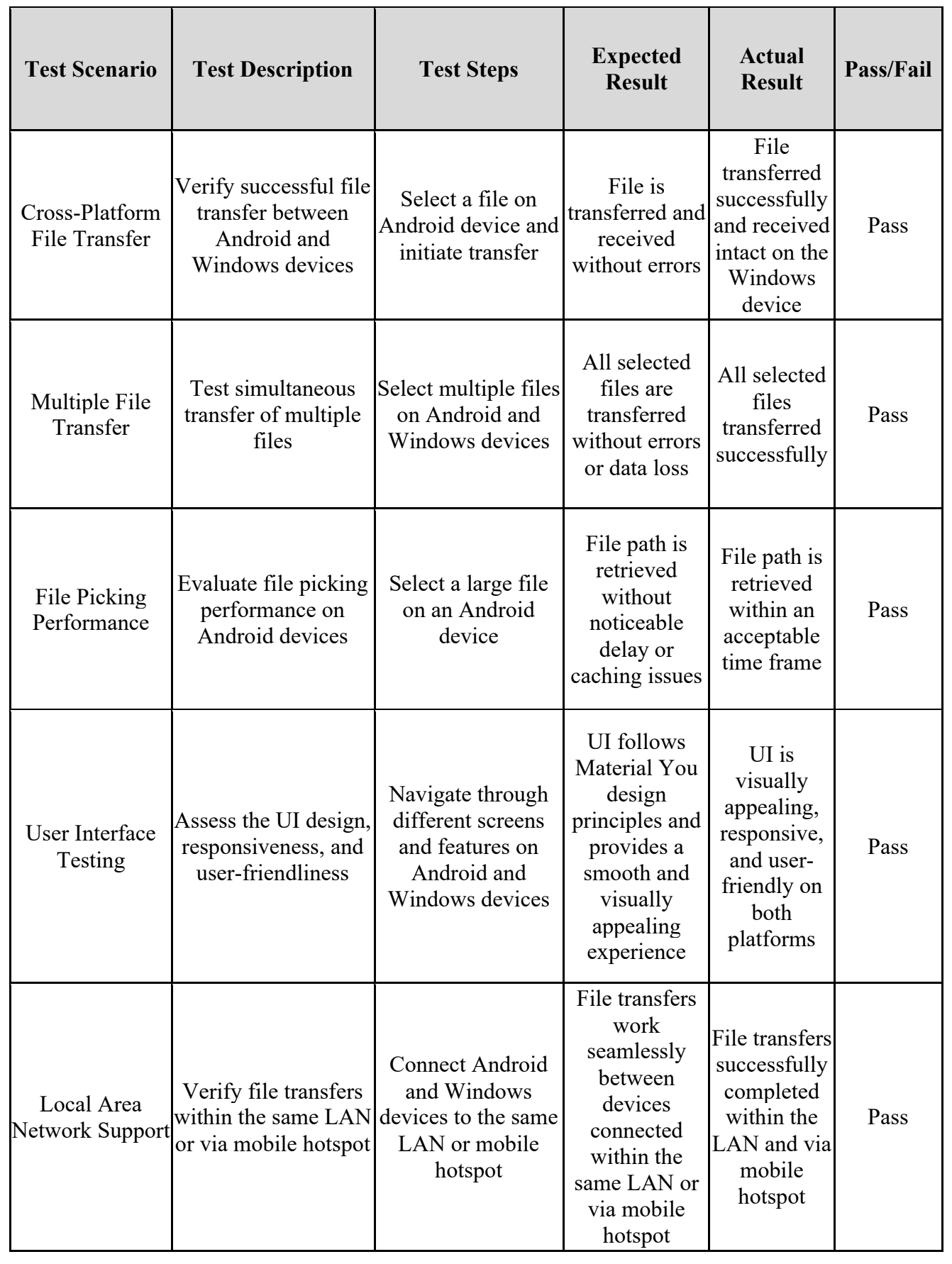}
\end{table}

\subsection{\textbf{CONCLUSION AND FUTURE ENHANCEMENTS}}\label{SCM}
A comprehensive solution to the problems with data transfer is provided by the suggested Peer- 
to-Peer Data Transfer App Within Local Area Network, which also delivers a quick, effective, and cross-platform experience. The programme enables direct communication between devices within a local area network by utilising a peer-to-peer network infrastructure, doing away with the requirement for a separate server. This strategy guarantees quicker data transfer speeds while simultaneously saving money. The programme optimises bandwidth usage because it does not rely on the internet, making it the best option for sending large files or sensitive data. 
Future updates can concentrate on extending support to Android TV devices to further improve the app. With this improvement, the app's functionality would be expanded and smooth data transfer between Android TV devices and other local network devices made possible. Additionally, a user-friendly and visually beautiful application will be made possible through ongoing improvements to the user interface and overall user experience.

\section*{\textbf{Acknowledgments}}

We would like to thank our institution "The National Institute of Engineering, Mysuru" For all the support rendered. Also we wish to acknowledge all the contributors of various technical papers and journals for their valuable contribution.

\section*{\textbf{References}}

\begin{enumerate}
    \item Wenhao Wu, “React Native vs Flutter, cross-platform mobile application frameworks”, Metropolia University of Applied Sciences Bachelor of Engineering Information technology Thesis 01 March 2018.\\
\url{https://www.theseus.fi/bitstream/handle/10024/146232/thesis.pdf?sequence=1} \\
    \item Paulo Roberto Martins de Andrade et al., “Cross Platform App: A Comparative Study”, University of Regina.\\
\url{https://www.researchgate.net/publication/271965071_Cross_Platform_App_A_Comparative_Study} \\
    \item Yoonsik Cheon et al., “Converting Android Native Apps to Flutter Cross-Platform Apps”, Las Vegas, NV, USA,2022.\\
\url{https://ieeexplore.ieee.org/document/9799091}\\

    \item Mr. D. S. Thosar et al., “A review on file sharing using LAN in peer-to-peer network”, S.V.I.T Chincholi, Nashik, 2021\\
    \item Suraj Bhul et al., “File Sharing Application: Turbo Share”, 2022.
\end{enumerate}

\end{document}